\documentclass[12pt]{iopart}
\usepackage{mathrsfs}
\usepackage{iopams}
\usepackage{bbding}
\usepackage{bm}
\begin{document}

\title{Protecting entangled states of two ions by engineering reservoir}
\author{Dong Xue, Jian Zou, Lin-Guang Yang, Jun-Gang Li and Bin Shao}
\address{Department of Physics, School of Science, and Key Laboratory of Cluster Science of Ministry of Education, Beijing Institute of Technology, Beijing 100081, People's Republic of China}
\ead{zoujian@bit.edu.cn}

\begin{abstract}

We present a proposal for realizing local decoherence-free evolution
of given entangled states of two two-level (TL) ions. For two TL
ions coupled to a single heavily damped cavity, we can use
engineering reservoir scheme to obtain a decoherence-free subspace
which can be nonadiabatically controlled by the system and reservoir
parameters. Then the local decoherence-free evolution of the
entangled states are achieved. And we also discuss the relation
between the geometric phases and the entanglement of the two ions
under the nonadiabatic coherent evolution.

\end{abstract}
\maketitle
\section{Introduction}

It is well known that one of the major obstacles for practical
manipulation of quantum information is the effect of decoherence due
to coupling to environment. So how to protect quantum information
from these effects become a major challenge. A number of strategies
have been proposed to fight decoherence, such as quantum error
correction \cite{P. W. Shor,A. M. Steane,A. Ekert}, decoherence-free
subspace (DFS) \cite{P. Zanardi,L. M. Duan,D. A. Lidar, D. Bacon},
dynamical decoupling (DD) \cite{G. S. Agarwa,L. Viola, D. Vitali},
engineering reservoir \cite{J. F. Poyatos}, \emph{etc}. The general
framework for DFS was introduced by Zanardi \emph{et al.} \cite{P.
Zanardi} in a spin-boson model undergoing both dephasing and
dissipation. In the presence of environment, DFS is a set of all
states which are immune to decoherence processes induced by
interaction with bath. These are groups of states that have robust
symmetry properties. The quantum information can be encoded in DFS
in order to increase reliability of quantum information processing
\cite{M. Mohseni,M. Bourennane,C. Langer}. On the other hand
Carvalho \emph{et al.} have presented a proposal for protecting
states of a trapped ion against decoherence, based on engineering of
pointer states \cite{A. R. R. Carvalho}. By controlling the
reservior, they have applied an indirect control on the protected
states, and have shown how to protect squeezed states, approximate
phase eigenstates, and superpositions of coherent states. Recently,
Prado \emph{et al.} have shown how to protect a nonstationary
superposition states of a two-level (TL) ion. Working with an ion
trapped inside a dissipative cavity they have constructed two
classes of decoherence-free evolution by broadening engineering
reservoir scheme \cite{F. O. Prado}. Under the assumption of a
squeezed engineered reservoir,  Carollo \emph{et al.} have proposed
a way to observe the adiabatic geometric phase acquired by a
protected state evolving coherently through the adiabatic
manipulation of the squeeze parameters of the engineered reservoir
\cite{A. Carollo}. Prado \emph{et al.} have generalized the method
proposed by  Carollo \emph{et al.} \cite{A. Carollo} and shown how
to observe the geometric phases acquired by the protected
nonstationary states even under the nonadiabatic evolution \cite{F.
O. Prado}. Authors in Refs. [15,16] only considered one particle
system, then a natural question arises, how about multi-particle
system? Can we use this engineering reservoir scheme to protect
entanglement? In this paper we will answer these questions.

In this paper we introduce a scheme, which generalizes  the
engineering reservoir scheme presented for a single ion in Ref.
\cite{F. O. Prado}, for protecting given entangled states of two TL
ions by building a class of periodic nonadiabatic coherent
evolution. We consider a system consisting of two TL ions coupled
resonantly to a dissipative cavity and simultaneously driven by
classical fields. For this system, we can obtain a time-independent
master equation by the techniques of engineering reservoir. Then we
can show that the given initial entangled state can be projected
into a DFS, and in particular, modifications of the parameters of
the reservoir may result in a controlled time evolution of the
protected subspace as a whole. In this way the protected entangled
state lying in this subspace evolves periodically. Finally, we
calculate the geometric phase of the whole system and the subsystems
under the nonadiabatic coherent evolution.

The paper is organized as follows. In Sec. 2, we introduce the
model and deduce the time-independent master equation by
 engineering reservoir. Then we obtain the DFS and show how to protect
 given entangled states of two TL ions. In Sec. 3, we calculate the geometric phase
of the whole system and the subsystems under the nonadiabatic
evolution and give the relation between the geometric phases and
the entanglement of the two TL ions. The conclusions are given in
Sec. 4.

\section{Model and results}
We consider a system consisting of a pair of two identical TL ions
$A$ and $B$ coupled resonantly to a single dissipative cavity mode
of frequency $\omega$ with equal coupling strength $g$. In this
paper we suppose that there are no direct interactions between the
two ions. The $i$th ion is driven by two classical fields of
frequencies $\omega^{i}_{l}$ with coupling strengths
$\Omega^{i}_{l}$ ($i=A, B$; $l=1,2$) where '$i$' and '$l$' represent
the ion and the classical field respectively. Within the
rotating-wave approximation and a sufficiently small Lamb- Dicke
parameter (to keep the motional state almost unchanged), the
Hamiltonian modeling the system is given by

\begin{equation} \label{eq:1}
\begin{array}{l}
H=\omega
a^{\dag}a+\frac{\omega_0}{2}(\sigma^{A}_{z}+\sigma^{B}_{z})+[(\sum_{l=1,2}\Omega^{A}_{l}e^{{\rm{i}}(\phi^{A}_{l}-\omega^{A}_{l}t)}\sigma^{A}_{eg}\\
\\
~~~~+\sum_{l=1,2}\Omega^{B}_{l}e^{{\rm{i}}(\phi^{B}_{l}-\omega^{B}_{l}t)}\sigma^{B}_{eg}+ga(\sigma^{A}_{eg}+\sigma^{B}_{eg}))+\rm{H.c.}],\\
\end{array}
\end{equation}
where $a^{\dag}$ ($a$) is the creation (annihilation) operator of
the cavity mode, $\omega_0$ is the transition frequency of the ions,
$\phi^{A}_{l}$ and $\phi^{B}_{l}$ are the dephasings of the
classical fields, and $\sigma^{i}_{kl}\equiv|k\rangle_{i}\langle
l|$, $k$ and $l$ being the ground state $g$ and excited state $e$ of
the ions. In the interaction picture, the Hamiltonian can be written
as:
\begin{equation} \label{eq:2}
\begin{array}{l}
H_1=[\sum_{l=1,2} (\Omega^{A}_{l}e^{\rm{i}\it
(\phi^{A}_{l}-\Delta^{A}_{l}t)}\sigma^{A}_{eg}+
\Omega^{B}_{l}e^{\rm{i}\it(\phi^{B}_{l}-\Delta^{B}_{l}t)}\sigma^{B}_{eg})\\
\\
~~~~~~+\rm{H.c.}] +\it[g e^{\rm{-i} \Delta t}a(\sigma^{A}_{eg}+\sigma^{B}_{eg})+\rm{H.c.}],\\
\end{array}
\end{equation}
where $\Delta=\omega-\omega_0$ and
$\Delta^{i}_{l}=\omega^{i}_{l}-\omega_0$. The first and second terms
of the Hamiltonian represent the classical field driving and the
coupling between the ions and the cavity respectively. We suppose
that the cavity mode is heavily damped with decay rate $\kappa$ and
the ions spontaneously decay with rates $\gamma^{i}$. Then the
master equation describing the system is given by
\begin{equation} \label{eq:1}
\dot \rho=- {\rm{i}}\left[ {H_1 ,\rho } \right] + \kappa D\left[ {a}
\right]\rho   + \sum_i\gamma^i D\left[ {\sigma^i_{ge}} \right]\rho,
 \end{equation}
 where the Lindblad decoherence
superoperator $\ D\left[a \right]\rho = a\rho a^\dag - a^\dag a\rho
/2 - \rho a^\dag a/2$ and $\ D\left[\sigma^i_{ge} \right]\rho  =
\sigma^i_{ge} \rho \sigma^{i}_{eg} - \sigma^{i}_{eg}
\sigma^i_{ge}\rho /2 - \rho \sigma^{i}_{eg} \sigma^i_{ge}/2$
describe cavity and ionic decays respectively.

Next we engineer the appropriate interaction between the ionic
levels and the cavity mode \cite{F. O. Prado}. Concretely, we adjust
the classical field $\omega^{i}_{1}$ in resonance with the ionic
transition frequency $\omega_0$, i.e., $\Delta^{i}_{1}=0$, and the
detuning of the field $\omega^{i}_{2}$ to
$\Delta^{i}_{2}=-2\Omega^{i}_{1}$. By applying the unitary
transformation $R(t)=R^{A}(t)\otimes R^{B}(t)$ to Hamiltonian (2),
where
$R^{i}(t)={\rm{exp}}[-{\rm{i}}\Omega^{i}_{1}t(e^{{\rm{i}}\phi^{i}_{1}}\sigma^{i}_{eg}+{\rm{H.c.}})]{\rm{exp}}\{-{\rm{i}}\Omega^{i}_{2}t[{\rm{cos}}(\varphi^{i})\sigma^{i}_{z}
+{\rm{i}}{\rm{sin}}(\varphi^{i})(e^{-{\rm{i}}\phi^{i}_{1}}\sigma^{i}_{ge}-e^{{\rm{i}}\phi^{i}_{1}}
\sigma^{i}_{eg})]\}$ with $\varphi^{i}=\phi^{i}_{1}-\phi^{i}_{2}$,
and adjusting
$\Omega^{i}_{1}=-\frac{1}{2}\Delta^{i}_{2}\gg\Omega^{i}_{2}=-\Delta\gg
g$, we can obtain the effective Hamiltonian in the rotating-wave
approximation
\begin{equation} \label{eq:3}
H_2=\frac{g}{2}[a^{\dag}(\sigma^{A}_{+-}+
\sigma^{B}_{+-})+a(\sigma^{A}_{-+}+\sigma^{B}_{-+})],
\end{equation}
where $|+\rangle_{i}={\rm{cos}}(\varphi^{i}/2)|e\rangle_{i}+
{\rm{i}}e^{-{\rm{i}}\phi^{i}_{1}}{\rm{sin}}(\varphi^{i}/2)
|g\rangle_{i}$ and
$|-\rangle_{i}={\rm{i}}e^{{\rm{i}}\phi^{i}_{1}}{\rm{sin}}(\varphi^{i}/2)|e\rangle_{i}+{\rm{cos}}(\varphi^{i}/2)
|g\rangle_{i}$.

Now applying the unitary transformation $R(t)$ to the original
master equation (3) we obtain
\begin{equation} \label{eq:4}
\dot{\rho^\prime}=-{\rm{i}}[H_2,\rho^\prime]+\kappa D\left[ {a}
\right]\rho^\prime+\sum_i\gamma^i D\left[ {\sigma^{\prime}}^{i}
\right]\rho^\prime,
\end{equation}
where $\rho^\prime=R(t)^\dag\rho R(t)$ and ${\sigma^{\prime}}^{i}
=R(t)^\dag\sigma^i_{ge} R(t)$. In the limit where the cavity decay
rate $\kappa$ is much larger than the other relevant frequencies,
the cavity mode can be adiabatically eliminated and we can obtain a
time-independent master equation just for the ions \cite{J. Wang}

\begin{equation} \label{eq:5}
\dot{\rho^\prime}=\Gamma D[\sigma^{A}_{+-}+
\sigma^{B}_{+-}]\rho^\prime+\sum_i\gamma^i D\left[
{\sigma^{\prime}}^{i} \right]\rho^\prime,
\end{equation}
where $\Gamma=g^{2}/\kappa$ is the coupling strength of the
engineered reservoir. Furthermore, under the assumption that
$\Gamma$ is much larger than the spontaneous emission rate
$\gamma^i$ of the ions, i.e., $\Gamma\gg\gamma^i$, Eq. (6) becomes

\begin{equation} \label{eq:6}
\dot{\rho^\prime}=\Gamma D[\sigma^{A}_{+-}+
\sigma^{B}_{+-}]\rho^\prime.
\end{equation}
From Eq. (7), the DFS is composed of all the eigenstates of the
operator $\sigma^{A}_{+-}+\sigma^{B}_{+-}$ with zero eigenvalue and
it is easy to prove that it is spanned by the following orthonormal
states: the product state
$|1\rangle\equiv|+\rangle_{A}|+\rangle_{B}$ and the maximal
entangled state
$|2\rangle\equiv\frac{1}{\sqrt{2}}(|-\rangle_{A}|+\rangle_{B}-|+\rangle_{A}|-\rangle_{B})$.
It is easy to see that the state
\begin{equation} \label{eq:7}
|\Psi_r\rangle=\sqrt{1-r}|1\rangle+\sqrt{r}~e^{{\rm{i}}\mu}|2\rangle,~(1\geq r \geq 0, \mu \in (0, 2\pi]),\\
\end{equation}
or a mixture state of $|1\rangle$ and $|2\rangle$ is the
equilibrium state of the master equation (7). If the two ions are
initially prepared in these states, they would remain the same.
However if the ions starts outside this subspace, the situation
gets more complicated. Because Eq. (7) is symmetric with respect
to exchange of the two ions, one could analyze the system in
terms of the antisymmetric
$$|2\rangle=\frac{(|-\rangle_{A}|+\rangle_{B}-|+\rangle_{A}|-\rangle_{B})}{\sqrt{2}}$$
and symmetric
$$|1\rangle=|+\rangle_{A}|+\rangle_{B},$$
$$|3\rangle=|-\rangle_{A}|-\rangle_{B},$$
$$|4\rangle=\frac{(|-\rangle_{A}|+\rangle_{B}+|+\rangle_{A}|-\rangle_{B})}{\sqrt{2}}.$$
subspaces \cite{A. R. R. Carvalho2}. Generally the initial state can
be decomposed into symmetric and antisymmetric components. The
symmetric part of the initial state evolves toward the state
$|1\rangle$ and the antisymmetric component $|2\rangle$ remain the
same.

In this paper, for simplicity we suppose that the protected state is
pure. Using concurrence as measure of degree of entanglement between
two qubits \cite{W. K. Wootters}, we can obtain that the
entanglement degree of the state $|\Psi_r\rangle$ is equal to $r$.
Since we obtain the master equation (7) through the unitary
transformation $R(t)$, the protected equilibrium state
$|\Psi_r\rangle$ is nonstationary in the original interaction
picture. Reversing the unitary transformation $R(t)$, the state
$|\Psi_r\rangle$ (for simplicity we take
$\Omega_1^A=\Omega_1^B=\Omega_1, \Omega_2^A=\Omega_2^B=\Omega_2$),
written in the interaction picture is
\begin{equation} \label{psi}
\begin{array}{l}
|\Psi(t)\rangle=R(t)|\Psi_r\rangle=c_1(t)|e\rangle_A|e\rangle_B+c_2(t)|e\rangle_A|g\rangle_B\\
\\
~~~~~~~~~~+c_3(t)|g\rangle_A|e\rangle_B+c_4(t)|g\rangle_A|g\rangle_B,\\
\end{array}
\end{equation}
where $c_1(t)$, $c_2(t)$, $c_3(t)$ and $c_4(t)$ are time-dependent
parameters:
\begin{equation} \label{eq:10}
\begin{array}{l}
c_1(t)=-\frac{{\rm{i}} e^{{\rm{i}} \mu }\sqrt{r}}{\sqrt{2}} [e^{{\rm{i}}\phi^B_1}\cos (\frac{\varphi^A}{2}-\Omega_1  t) \sin (\frac{\varphi^B}{2}-\Omega_1  t)\\
~~~~~~~~~~-e^{{\rm{i}} \phi^A _1}\sin (\frac{\varphi^A}{2}-\Omega_1  t) \cos (\frac{\varphi^B}{2}-\Omega_1  t) ]\\
~~~~~~~~~~+ e^{-{\rm{i}} 2 \Omega_2  t }\sqrt{1-r} \cos (\frac{\varphi^A}{2}-\Omega_1  t) \cos (\frac{\varphi^B}{2}-\Omega_1  t),\\
c_2(t)=-\frac{ e^{{\rm{i}} \mu }\sqrt{r}}{\sqrt{2}}[  \cos (\frac{\varphi^A}{2}-\Omega_1  t) \cos (\frac{\varphi^B}{2}-\Omega_1  t)\\
~~~~~~~~~~+e^{{\rm{i}} (\phi^A_1-\phi^B_1)} \sin (\frac{\varphi^A}{2}-\Omega_1  t) \sin (\frac{\varphi^B}{2}-\Omega_1  t)]\\
~~~~~~~~~~+{\rm{i}} e^{-{\rm{i}}( \phi^B_1+2 \Omega_2  t)} \sqrt{1-r} \cos (\frac{\varphi^A}{2}-\Omega_1  t) \sin (\frac{\varphi^B}{2}-\Omega_1  t),\\
c_3(t)=\frac{ e^{{\rm{i}} \mu }\sqrt{r}}{\sqrt{2}}[\cos (\frac{\varphi^A}{2}-\Omega_1  t) \cos (\frac{\varphi^B}{2}-\Omega_1  t)\\
~~~~~~~~~+e^{{\rm{i}} (\phi^B_1-\phi^A_1)}  \sin (\frac{\varphi^A}{2}-\Omega_1  t) \sin (\frac{\varphi^B}{2}-\Omega_1  t)]\\
~~~~~~~~~+{\rm{i}} e^{-{\rm{i}} (\phi^A_1+2 \Omega_2  t)} \sqrt{1-r}\sin (\frac{\varphi^A}{2}-\Omega_1  t) \cos (\frac{\varphi^B}{2}-\Omega_1  t),\\
c_4(t)=-\frac{{\rm{i}} e^{{\rm{i}} \mu }\sqrt{r}}{\sqrt{2}}[e^{-{\rm{i}} \phi^A_1} \sin (\frac{\varphi^A}{2}-\Omega_1  t) \cos (\frac{\varphi^B}{2}-\Omega_1  t)\\
~~~~~~~~~-e^{-{\rm{i}} \phi^B_1} \cos (\frac{\varphi^A}{2}-\Omega_1  t) \sin (\frac{\varphi^B}{2}-\Omega_1  t)]\\
~~~~~~~~~-e^{-{\rm{i}} (\phi^A_1+\phi^B_1+2 \Omega_2  t)} \sqrt{1-r}\sin (\frac{\varphi^A}{2}-\Omega_1  t) \sin (\frac{\varphi^B}{2}-\Omega_1  t).\\
\end{array}
\end{equation}
The $\Omega_1$, $\Omega_2$, $\phi^{i}_{1}$, and
$\varphi^{i}=\phi^{i}_{1}-\phi^{i}_{2}$ are adjustable parameters of
the classical fields. The nonstationary protected state
$|\Psi(t)\rangle$ is allowed for a nonadiabatic coherent evolution,
which can be manipulated through those parameters $\Omega_1$,
$\Omega_2$, $\phi^{i}_{1}$ and $\varphi^{i}$. Under the assumption
that $\Omega_1$=$N\Omega_2$ (where $N$ is an integer and $N\gg1$),
the evolution is periodic and the period is equal to $\pi/\Omega_2$.
That is to say, we can obtain the protected initial state at time
$n\pi/\Omega_2$ $(n=1,2,\cdots)$ and the concurrence of the system
is invariable in the evolution. Concretely, suppose the initial
state be $
|\Psi(0)\rangle=c_1(0)|e\rangle_A|e\rangle_B+c_2(0)|e\rangle_A|g\rangle_B+c_3(0)|g\rangle_A|e\rangle_B+c_4(0)|g\rangle_A|g\rangle_B,$
where $c_{i}(0)$ are known complex constants and
$\sum_{i}|c_{i}(0)|^{2}=1$ $(i=1,2,3,4$). Let $t=0$ in  Eq. (10), we
can obtain 4 equations, and then we can obtain the required
parameters of the engineering reservoir $\phi^{i}_{1}$ and
$\varphi^{i}$ and the coefficients $\mu$ and $r$ of the
corresponding state $|\Psi_r\rangle$ in the DFS, i.e., we can choose
these parameters $\phi^{i}_{1}$ and $\varphi^{i}$
 to project the given initial state $|\Psi(0)\rangle$  to the corresponding state $|\Psi_r\rangle$ in
the DFS. As a simple example, we suppose that the initial state is a
maximal entangled state $|\Psi(0)\rangle$ (Bell states). From Eqs.
(9) and (10), we obtain the parameters of the engineering reservoir
and the corresponding protected state $|\Psi_r\rangle$ in the DFS
(in table 1).
\begin{table}
\caption{\label{arttype}parameters of the engineering reservoir and
the corresponding state $|\Psi_r\rangle$.}
 \footnotesize\rm
\begin{tabular*}{\textwidth}{@{}l*{15}{@{\extracolsep{0pt plus12pt}}l}}
\br
initial state $|\Psi(0)\rangle$ & parameters of the engineering reservoir & corresponding state $|\Psi_r\rangle$ in DFS\\
\mr
$\frac{1}{\sqrt{2}}(|e\rangle_A|e\rangle_B+|g\rangle_A|g\rangle_B)$ & $\phi^{A} _1=\frac{\pi}{2},\phi^{B}_1=\frac{\pi}{2}, \varphi^{A}=0, \varphi^{B}=\pi$& $\frac{1}{\sqrt{2}}(|-\rangle_{A}|+\rangle_{B}-|+\rangle_{A}|-\rangle_{B})$\\
$\frac{1}{\sqrt{2}}(|e\rangle_A|e\rangle_B-|g\rangle_A|g\rangle_B) $& $\phi^{A}_1=0, \phi^{B}_1=0,\varphi^{A}=0,\varphi^{B}=\pi$& $ \frac{\rm{i}}{\sqrt{2}}(|-\rangle_{A}|+\rangle_{B}-|+\rangle_{A}|-\rangle_{B}) $ \\
$\frac{1}{\sqrt{2}}(|e\rangle_A|g\rangle_B+|g\rangle_A|e\rangle_B)$ & $\phi^{A} _1=-\frac{\pi}{2},\phi^{B}_1=0,\varphi^{A}=\pi, \varphi^{B}=\pi$&$\frac{-\rm{i}}{\sqrt{2}}(|-\rangle_{A}|+\rangle_{B}-|+\rangle_{A}|-\rangle_{B})$\\
$\frac{1}{\sqrt{2}}(|e\rangle_A|g\rangle_B-|g\rangle_A|e\rangle_B)$ & $\phi^{A}_1=-\frac{\pi}{2},\phi^{B}_1=-\frac{\pi}{2},~\varphi^{A}=\pi, \varphi^{B}=\pi~~~$& $\frac{-1}{\sqrt{2}}(|-\rangle_{A}|+\rangle_{B}-|+\rangle_{A}|-\rangle_{B})$\\
\br
\end{tabular*}
\end{table}
Now we consider a more general example,
$|\Psi(0)\rangle=m|e\rangle_A|e\rangle_B+n
e^{\rm{i}\theta}|g\rangle_A|g\rangle_B$, where $m,n \geq 0$
satisfying $m^{2}+n^{2}=1$, and $\theta \in (0,2\pi]$. From Eqs. (9)
and (10), we can obtain the parameters of the classical fields
\begin{equation} \label{eq:9}
\begin{array}{l}
 ~\phi^{A}_1=\phi^{B}_1=\frac{1}{2}(\pi -\theta ),\\
\left.
\begin{array}{l}
 \varphi^A~=\frac{\pi}{2}[1+{\rm{sign}}(n -m )]
           -{\rm{arctan}}\left(\frac{2 \sqrt{m n }}{|m-n|}\right),\\
 \varphi ^B=\frac{\pi}{2}[1+{\rm{sign}}(n -m )]
           +{\rm{arctan}}\left(\frac{2 \sqrt{m n
           }}{|m-n|}\right),
 \end{array}
 \right\}(m \neq n)
\end{array}
\end{equation}
and the corresponding state $|\Psi_r\rangle=\sqrt{m
n}~e^{\rm{i}\frac{\theta}{2}}(|-\rangle_{A}|+\rangle_{B}-|+\rangle_{A}|-\rangle_{B})+\sqrt{1-2
m n}~|+\rangle_{A}|+\rangle_{B}$ in the DFS. It must be noted that
although many methods have been proposed to protect the entanglement
from dissipation, our approach is different. Most of the schemes to
protect entanglement is static, but ours is dynamic. More
specifically the protected state goes through a cyclic nonadiabatic
coherent evolution, but the degree of entanglement does not evolves
and remains the same. In this way we can change one entangled state
into another entangled state with the same degree of entanglement
against dissipation.

All our discussions were based on Eq. (7), where spontaneous
emission effects were neglected. However, spontaneous emission is
the fundamental limiting factor for the existence of entanglement in
a system of ions. The Eq. (6) describes the ionic system including
the effect of spontaneous emission which introduces a coupling
between the symmetric and antisymmetric subspaces. Next, we will
analyze the spontaneous emission effects in our protected schemes.
As a example, we consider that the two TL ions are initially
prepared in
\begin{equation}\label{eq:18}
|\Psi_E\rangle=\frac{1}{\sqrt{2}}|g\rangle_{A}|g\rangle_{B}+
\frac{1}{2}(|e\rangle_{A}|g\rangle_{B}-|g\rangle_{A}|e\rangle_{B}),
\end{equation}
Without spontaneous decay, from Eqs. (\ref{psi}) and (10), we can
see that by adjusting the parameters of the classical fields
\begin{equation} \label{eq:19}
 \varphi^A~=\varphi^B=\pi,~ \phi^A_1=0,~\phi^B_1=\pi,~(\rm{or}~~\phi^A_1=\pi,~\phi^B_1=0),
\end{equation}
the initial state $|\Psi_E\rangle$ will undergo a coherent local
evolution and the $|\Psi_E(t)\rangle$ can be written as
\begin{equation} \label{eq:20}
\begin{array}{l}
|\Psi_E(t)\rangle=(\frac{e^{-2{\rm{i}}  \Omega_2t} {\rm{sin}}[\Omega_1t]^2}{\sqrt{2}}-\frac{{\rm{i}}{\rm{sin}}[2 \Omega_1t]}{2})|e\rangle_A|e\rangle_B\\
~~~~~~~~~~+(\frac{{\rm{cos}}[2  \Omega_1t]}{2} +\frac{{\rm{i}} e^{-2 {\rm{i}}  \Omega_2t} {\rm{sin}}[2 \Omega_1t]}{2\sqrt{2}})|e\rangle_A|g\rangle_B\\
~~~~~~~~~~-(\frac{{\rm{cos}}[2  \Omega_1t]}{2} +\frac{{\rm{i}} e^{-2 {\rm{i}} \Omega_2t} {\rm{sin}}[2 \Omega_1t]}{2\sqrt{2}})|g\rangle_A|e\rangle_B\\
~~~~~~~~~~+(\frac{e^{-2{\rm{i}}  \Omega_2t} \rm{cos}[ \Omega_1t]^2}{\sqrt{2}}+\frac{{\rm{i}}{\rm{sin}}[2 \Omega_1t]}{2})|g\rangle_A|g\rangle_B.\\
\end{array}
\end{equation}
Now instead of Eq.(7) we  numerically solve Eq.(6), which include
the effect of spontaneous emission for the given initial state
$|\Psi_E\rangle$. We can compute the fidelity $F=\rm
{Tr}[|\Psi_E(t)\rangle\langle\Psi_E(t)|R(t)\rho^\prime_E(t)R^{\dag}(t)]$,
where $\rho^\prime_E(t)$ is the solution of Eq.(6). Within the
regime $\Omega_1$=10$\Omega_2$=100$g$, $g$=500$\gamma^A$,
$\gamma^A$=$\gamma^B$ and $\kappa$=3$g$, the fidelity $F$ as a
function of time is shown in Fig. 1. From Fig. 1, we can see that
the spontaneous emissions move the system away from the protected
state $|\Psi_E(t)\rangle$. If the spontaneous emission rate is
very small, the fidelity reduces slowly, for example, when
$t$=$100\pi/{\Omega_2}$, the fidelity is around 96.9$\%$.
Therefore, as long as the spontaneous emission rate $\gamma_i$ is
much smaller than the other relevant frequencies of the problem,
we can neglect the spontaneous emission effect within a finite
time, such as $t\ll 100\pi/{\Omega_2}$ in Fig.1. It is worth
stressing that our protected scheme might be realized
experimentally. The setup of two atoms equally coupled to a cavity
mode with possibility of individual addressing has already been
demonstrated in \cite{S. Numann}. The large cooperativity
parameter ( $\Gamma=g^{2}/\kappa\gg\gamma$) has been obtained in a
variety of recent experiments \cite{C. J. Hood,A. D. Boozer,P.
Maunz}.

\section{Relation between entanglement and geometric phase of the system}

From above discussion, we can see that the protected two-ion state
(9) is 'dynamic' rather than 'static'. If $\Omega_1$=$N\Omega_2$
(where $N$ is an integer and $N\gg1$), the system will undergo a
cyclic coherent evolution. After a cyclic evolution, the system
returns to its original state but may acquire a geometric phase. To
compute the geometric phase we use the definition given in Ref.
\cite{N. Mukunda}. From Eq. (\ref{psi}), after a cyclic evolution,
$\tau=\pi/\Omega_2$, the acquired geometric phase is
\begin{equation}\label{BP}
\beta^{G}(\tau)={\rm{i}}\int_{0}^{\tau}\langle\Psi(t)|\frac{d}{dt}|\Psi(t)\rangle
dt=2\pi(1-r).\\
\end{equation}
From Eq. (15) it  can be seen that, the geometric phase of the
whole system is only a simple linear function of the entanglement
degree $r$, and have nothing to do with other system parameters.
Next we calculate the geometric phase of the subsystems and to
study the relation between the geometric phase of the subsystems
and the entanglement degree. Generally speaking, the state of the
subsystem is no longer a pure one, so we adopt the definition of
geometric phase for mixed states under bilocal unitary evolution
\cite{D. M. Tong}. If the Schmidt coefficients are nondegenerate,
after a cyclic evolution, the geometric phase of the subsystem can
be written as
 \begin{equation}\label{GP}
 \begin{array}{l}
 \beta^{i}(\tau)=\arg[\sum_{k=1}^N p_k\langle\mu_k|U^{i}(\tau)|\mu_k\rangle\\
 \\
 ~~~~~~~~~~\times {\rm{exp}} (-\int_0^\tau\langle\mu_k|U^{i
 \dag}(t)\dot{U}^{i}(t)|\mu_k\rangle dt)],
 \end{array}
 \end{equation}
 where $p_k$ is the Schmidt coefficient, $|\mu_k\rangle$ is the
 corresponding eigenstate of the reduced density matrix $\rho^{i}$ (obtained after tracing over the
 other ion) and $U^{i}(t)$ is a local unitary evolution operator acting on
 the $i$-th ion.

For our system, if the initial state is not maximally entangled
($r\neq1$), the Schmidt coefficients are nondegenerate. Using Eq.
(16), we can obtain the geometric phase of the subsystems under
bilocal unitary evolution $R(t)$
\begin{equation}\label{subgp}
\beta^A=\beta^B={\rm{arg}}[\cos (
\sqrt{\frac{1-r}{1+r}}\pi)+{\rm{i}} \sqrt{1-r^2}\sin
(\sqrt{\frac{1-r}{1+r}}\pi)]  \\
\end{equation}
When the initial state is a maximal entangled state, i.e., $r=1$,
the Schmidt coefficients are degenerate and the reduced density
matrix of the subsystem at $t=0$ is
$$\rho^{A(B)}(0)=\frac{1}{2}I.$$
Because the system subjects to the bilocal unitary evolution $R(t)$,
the reduced density matrix of the subsystem at any time $t$ is
\begin{equation}
\begin{array}{l}
\rho^{A(B)}(t)={\rm{Tr}_{B(A)}}[R^{A}(t)\otimes R^{B}(t) \rho(0)
R^{A\dag}(t)\otimes R^{B\dag}(t)] \\
\\
~~~~~~~~~~~=\frac{1}{2}I,\\
\end{array}
\end{equation}
which means
\begin{equation}
\beta^{A(B)}=0.
\end{equation}
 From Eqs. (17) and (19), we can obtain the relation between the geometric phase of the subsystems and the
entanglement degree. Again the geometric phase of the subsystems
$\beta^{A(B)}$ is also only a function of the entanglement degree
$r$ which can be seen from Eq. (17) and is shown in Fig. 2. It can
be seen form Fig. 2 that $\beta^{A(B)}$ is a monotonic decreasing
function of $r$. If we could measure the geometric phase of the
subsystems $\beta^{A(B)}$, we can infer the entanglement degree
$r$ of the protected entangled state.

\section{Conclusion}
In this paper, we have considered two TL ions in a heavily damped
cavity. Using engineering reservoir scheme, we have obtained a
time-independent master equation, and then have found a DFS for
this master equation, which can be nonadiabatically controlled by
the system-reservoir parameters. We have achieved a class of
decoherence-free cyclic evolution of the entangled state. Finally,
we have calculated the geometric phases of the whole system and
the subsystems under the nonadiabatic coherent evolution, and have
found that there is one-to-one correspondence between the
geometric phase of the whole system, the geometric phase of
subsystems and the entanglement degree $r$ .

\section*{Acknowledgment}
This work was supported by National Natural Science Foundation of
China (Grants No. 10974016, No. 11005008, and No. 11075013).

\newpage

\begin{center}
\large\bf CAPTIONS \normalsize\rm
\end{center}

Fig.1 The fidelity F as a function of time.

Fig.2 The geometric phase of the subsystems $\beta^{A(B)}$ as a
function of entanglement degree $r$.


\section*{References}
\begin{thebibliography}{21}
\bibitem{P. W. Shor} Shor P W 1995 \emph{Phys. Rev. A} \textbf{52} R2493
\bibitem{A. M. Steane}  Steane A M 1996 \emph{Phys. Rev. Lett.} \textbf{77} 793
\bibitem{A. Ekert} Ekert A and Macchiavello C 1996 \emph{Phys. Rev. Lett.} \textbf{77} 2585
\bibitem{P. Zanardi} Zanardi P and Rasetti M 1997 \emph{Phys. Rev. Lett. } \textbf{79} 3306
\bibitem{L. M. Duan}Duan L M  and Guo G C 1998 \emph{Phys. Rev. A} \textbf{57} 737
\bibitem{D. A. Lidar} Lidar D A, Chuang I L and Whaley K B 1998 \emph{Phys. Rev. Lett.} \textbf{81} 2594
\bibitem{D. Bacon} Bacon D, Kempe J, Lidar D A and Whaley K B 2000 \emph{Phys. Rev. Lett.} \textbf{85} 1758
\bibitem{G. S. Agarwa}Agarwal G S 1999 \emph{Phys. Rev. A} \textbf{61} 013809
\bibitem{L. Viola} Viola L, Knill E and Lloyd S 1999 \emph{Phys. Rev. Lett.} \textbf{82} 2417
\bibitem{D. Vitali} Vitali D and Tombesi P 1999 \emph{Phys. Rev. A} \textbf{59} 4178
\bibitem{J. F. Poyatos} Poyatos J F, Cirac J I and Zoller P 1996 \emph{Phys. Rev. Lett.} \textbf{77} 4728
\bibitem{M. Mohseni}Mohseni M, Lundeen J S, Resch K J and Steinberg A M 2003  \emph{Phys. Rev. Lett.} \textbf{91} 187903
\bibitem{M. Bourennane}Bourennane M, Eibl M, Gaertner S, Kurtsiefer C, Cabello A and Weinfurter H 2004  \emph{Phys. Rev. Lett.} \textbf{92} 107901
\bibitem{C. Langer} Langer C,  Ozeri R, Jost J D,  Chiaverini J, DeMarco B, Ben-Kish A, Blakestad R B, Britton J, Hume D B, Itano W M, Leibfried D, Reichle R, Rosenband T, Schaetz T, Schmidt P O and Wineland D J 2005  \emph{Phys. Rev. Lett.} \textbf{95} 060502
\bibitem{A. R. R. Carvalho}Carvalho A R R, Milman P, de Matos Filho R L and Davidovich L 2001 \emph{Phys. Rev. Lett.} \textbf{86} 4988
\bibitem{F. O. Prado} Prado F O, Duzzioni E I, Moussa M H Y, de Almeida N G and Villas-B\^{o}as C J 2009 \emph{Phys. Rev. Lett.} \textbf{102} 073008
\bibitem{A. Carollo} Carollo A, Paternostro G M, {\L}ozinski A, Santos M F, and Vedral V 2006 \emph{Phys. Rev. Lett.} \textbf{96} 150403
\bibitem{J. Wang} Wang J, Wiseman H M and Milburn G J 2005 \emph{Phys. Rev. A} \textbf{71} 042309
\bibitem{A. R. R. Carvalho2} Carvalho A R R, Reid A J S and Hope J J 2008 \emph{Phys. Rev. A} \textbf{78} 012334
\bibitem{W. K. Wootters} Wootters W K 1998 \emph{Phys. Rev. Lett.} \textbf{80} 2245
\bibitem{S. Numann} Nu{\ss}mann S, Hijlkema M, Weber B, Rohde F, Rempe G and Kuhn A 2005 \emph{Phys. Rev. Lett.} \textbf{95} 173602
\bibitem{C. J. Hood} Hood C J, Kimble H J and Ye J 2001 \emph{Phys. Rev. A} \textbf{64} 033804
\bibitem{A. D. Boozer} Boozer A D, Boca A, Miller R, Northup T E and Kimble H J 2006 \emph{Phys. Rev. Lett.} \textbf{97} 083602
\bibitem{P. Maunz} Maunz P, Puppe T, Schuster I, Syassen N, Pinkse P W H and Rempe G 2005 \emph{Phys. Rev. Lett.} \textbf{94} 033002
\bibitem{N. Mukunda} Mukunda N and Simon R 1993 \emph{Ann. Phys. (Leipzig)} \textbf{228} 205
\bibitem{D. M. Tong} Tong D M, Sj\"{o}qvist E, Kwek L C, Oh C H and Ericsson M 2003 \emph{Phys. Rev. A} \textbf{68} 022106

\end{thebibliography}
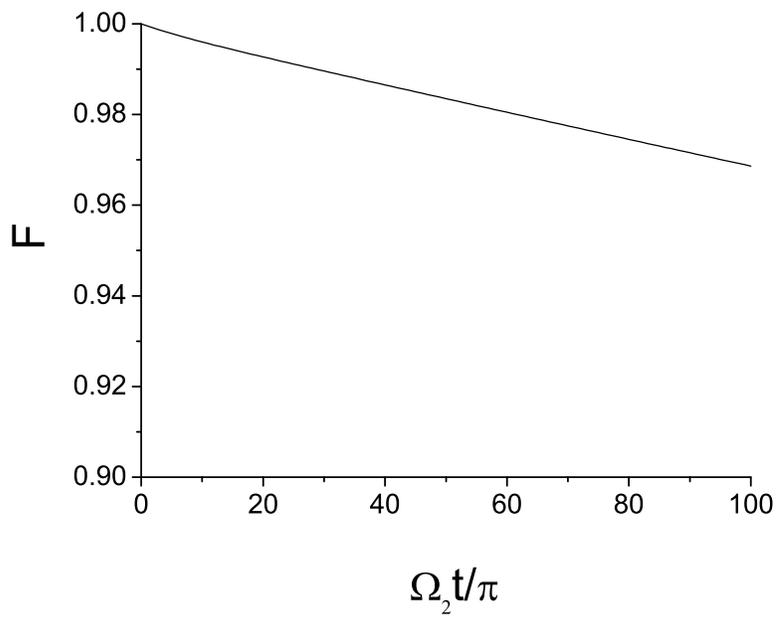
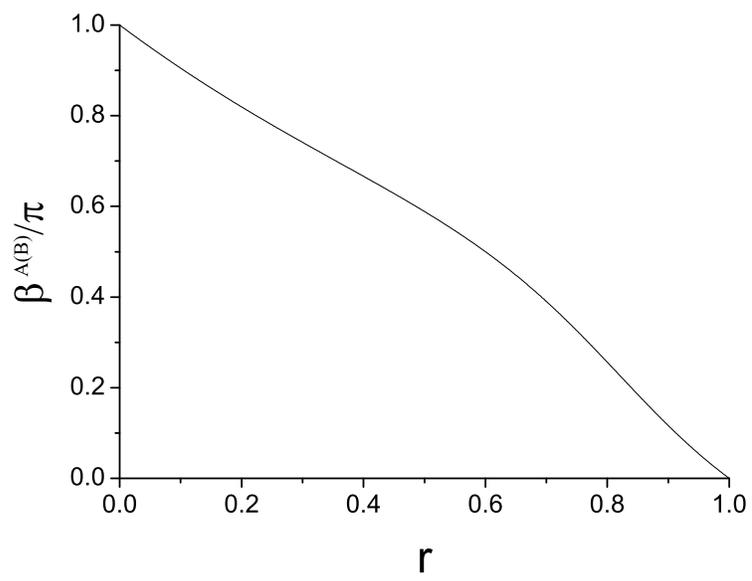
\end{document}